\documentclass[prl,10pt,twocolumn,showpacs]{revtex4}

\usepackage{dcolumn,bm,amsmath,graphicx}

\begin{document}

\title{Immunization Dynamics on a 2-layer Network Model}

\author{Hang-Hyun Jo}
\email[e-mail: ]{kyauou2@kaist.ac.kr}
\author{Hie-Tae Moon}
\author{Seung Ki Baek}

\affiliation{Department of Physics, Korea Advanced Institute of Science and Technology, Daejeon 305-701, Republic of Korea}

\begin{abstract}
We introduce a 2-layer network model for the study of the immunization dynamics in epidemics. Spreading of an epidemic is modeled as an excitatory process in a small-world network (body layer) while immunization by prevention for the disease as a dynamic process in a scale-free network (head layer). It is shown that prevention indeed turns periodic rages of an epidemic into small fluctuation. The study also reveals that, in a certain situation, prevention actually plays an adverse role and helps the disease survive. We argue that the presence of two different characteristic time scales contributes to the immunization dynamics observed. 
\end{abstract}

\maketitle

Recently the agent of severe acute respiratory syndrome (SARS), a novel coronavirus, has spread over Asia. The epidemic apparently originated in the Guandong province of the People's Republic of China, and then spread rapidly throughout the world via air travel \cite{Riley2003}. For physicists the dynamics of epidemic spreading and immunization has been one of the fairly intriguing subjects. An amount of epidemiological work has been done on various networks; epidemic models of an infectious disease on small-world networks \cite{KA2001} and computer virus infections on both original and structured scale-free networks \cite{PV2001,EK2002}, which seems to be due to the topology of Internet, etc. The effect of immunization is also studied, which shows that the targeted immunization strategy is more effective than the random immunization to lower the vulnerability in complex networks \cite{PV2002,ZK2002}. In these previous Refs. the effect of immunization on selected nodes is permanent and initially vaccinated nodes cannot infect or be infected by others in the future so that they function as blocked passage or are merely removed from the viewpoint of the disease. Therefore this immunization by prevention is concerned with the static network structure rather than some dynamical properties of models. For the more realistic prevention models let us consider the prevention as dynamical as the disease.

Various information, which covers all the interactions among the elements of a system, spreads through the complex network. In a social network, one usually obtains many kinds of information through more than one channel, and which one is employed depends on the characteristics of the information, e.g. local interactions by physical contacts vs. global ones by mass media or Internet, public ones by way of official announcements vs. private ones at the club, etc. One can separate several kinds of channels from the whole network and make multi-layered network model, which means that the different kinds of information propagate through the different layers and interact with each other when they meet at a node. The hybridization of networks was already attempted in Ref. \cite{NWmodel}, indicating that the regular lattice superposed with the sparse random graph shows the clustering among the nodes and the small-world effect simultaneously. Here we propose a 2-layer network model composed of a small-world network and a scale-free network, then analyze how the different kinds of information, which are the disease spreading and prevention in this study, compete with each other.

Small-world and scale-free networks are the most frequently observed types, and which one is more proper again depends on what we are concerned about. For example, it can be a scale-free network in case of a venereal disease since it is found that the number of sexual partners shows a scale-free distribution \cite{LEA2001}. Generally speaking, however, interactions between individuals are limited by physical restriction such as spatial localization and the cost of adding social links \cite{Amaral2000}, henceforth most social networks would be modeled by the single-scale network, e.g. Watts-Strogatz small-world network, or multi-scale network rather than a scale-free one. On the other hand, we expect a scale-free network to be suitable for spreading news or for governing the whole system owing to the existence of hubs and hierarchical structure. Since the prevention does not usually occur by physical contacts but most by means of communicating media or by some health policy, we separate a prevention strategy using a scale-free network, such as Internet, where an infected individual is supposed to report the status to his or her on-line nearest neighbors. Thus the existing prevention strategies should be modified into scale-free ones. We investigate a seemingly common situation where we control the small-world excitatory system through scale-free inhibitory system. 

\begin{figure}
\includegraphics[angle=0,scale=.5]{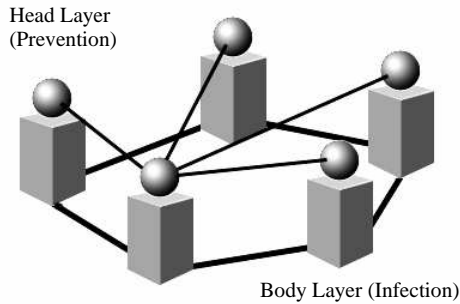}
\caption{The epidemic spreading and immunization model with 2-layer structure. The infectious disease spreads over the Body Layer while the information for prevention does over the Head Layer. A susceptible node evolves to get infected or immune by its infected neighbors both on the HL and the BL.}
\end{figure}

We prepare a ground where an epidemic and immunization by prevention directly compete with each other, so that we may get more information on the dynamics. For that purpose we introduce a 2-layer epidemic spreading and immunization model, or briefly `2-layer model'. Figure 1 displays the schematic structure of the model; the lower layer (body layer; BL) is a physical network over which the infectious disease spreads, while the upper one (head layer; HL) is an information network used for immunization by prevention. We notice that each channel has a different connectivity structure. We adopt here the well known susceptible-infected-refractory ($SIR$) model \cite{Murray1993} for the characteristic states of a disease so that the health of each node is described by one of $S$, $I$, and $R$. An $S$-state node, when infected through contagion, starts an infection cycle of disease, i.e. it passes to the $I$-state, where it remains for a certain infection period $\tau_{I}$ and then to a recovery period $\tau_{R}$ before returning to the $S$-state. If an $S$-state node gets immune through the information channel, it stays for a prevention period $\tau_{P}$, then returns to the $S$-state, which is a prevention cycle (Fig. 2). An individual can get infected only during the $S$-state and only by contacting $I$-state neighbors, while it neither infects nor is infected by others during the $R$-state. In this sense $R$-state nodes in both cycles do the same function as immune ones. Therefore it is necessary to divide the immunization into two classes; one is through infection and the other is by prevention. The existing prevention strategies correspond to the case of assigning the infinite $\tau_{P}$ for initially selected nodes and with no further immunization. For a dynamical immunization we set both $\tau_{R}$ and $\tau_{P}$ to be finite and there is no initial selection of immune nodes. If the $S$-state nodes are immunized only by the immediate inoculation, $\tau_{R}$ and $\tau_{P}$ must be the same. Now we complete the 2-layer $SIRS$ spreading and $SRS$ prevention model. 

\begin{figure}
\includegraphics[angle=0,scale=.45]{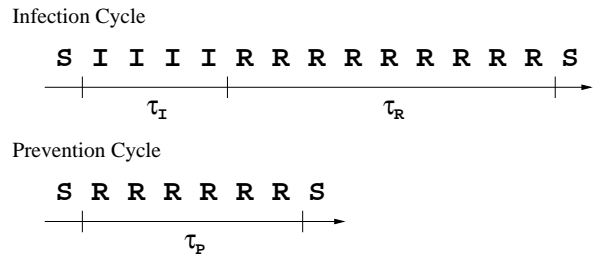}
\caption{The infection and prevention cycle of a node. At once a susceptible node becomes infected it goes through the infection cycle deterministically. When it has $I$-state it can contaminate its neighbors. After the infection period it becomes refractory and during the recovery period it becomes immune but not contagious. Prevention by its infected neighbors lasts for a prevention period. If nodes are immunized by a biological method $\tau_{R}$ and $\tau_{P}$ must be the same.}
\end{figure}

In the original small-world model, the rewiring probability $p$ characterizes the degree of disorder, ranging from a regular lattice to a completely random graph \cite{WS1998,Watts1999}, and is defined here as the number of randomly rewired links divided by the total number of links of the initial regular lattice. Hence at $p=0$, $N$ nodes constitutes a simple regular lattice each node of which has $k_0$ nearest neighbors, and becomes more random as $p$ increases. We choose $k_0$ large enough to guarantee that the network shall be connected even after many random rewirings. Even at low $p$, the network shows small-world effect, i.e. the characteristic path length $L(p)$, the averaged minimum path length between two randomly chosen nodes, increases logarithmically with the system size, while $L(p)$ does linearly in case of the purely regular lattice. On the other hand, a scale-free network is constructed by two rules: growth and preferential attachment \cite{BA1999,BA2000}. Starting with a small number ($m_{0}$) of nodes, we add a new node with $m$ links at every time step. Then the probability that a new node is connected to node $i$ is directly proportional to $k_i$, the number of neighbors of node $i$. The resulting probability distribution of the node connectivity, $P(k)$, follows a power law: $P(k) \sim  k^{-3}$. The power law implies rather frequent emergence of highly connected nodes, often called hubs, which drastically reduce path lengths.

Let us define necessary parameters for the model study. A node with $S$-state may have $k_{b}$ infected neighbors, each of which has probability $q_{b}$ of contagion. At the same time, the act of prevention triggered by $k_{h}$ infected neighbors, each of which has the probability $q_{h}$ of immunization by prevention, protects a node from being infected for a prevention period. With these provisions, a node is expected to evolve according to the following probabilities:
\begin{equation}
\begin{split}
&P(S\rightarrow S) =  (1-q_{h})^{k_{h}} (1-q_{b})^{k_{b}}, \\
&P(S\rightarrow I) =  (1-q_{h})^{k_{h}} \{1-(1-q_{b})^{k_{b}}\}, \\
&P(S\rightarrow R) =  1-(1-q_{h})^{k_{h}}.
\end{split}
\end{equation}

Starting with $m_{0}=m=3$, we construct HL as a Barab\'asi-Albert scale-free network with $N=10^{4}$ and BL as a Watts-Strogatz small-world network with $N=10^{4}$, $k_0=6$, and $p=0.5$. The infection period and the recovery period ($=$ the prevention period) are set to be $4$ and $9$ time steps, respectively, and initial infected fraction is to be $0.1$. These values are selected following Ref. \cite{KA2001}. Other initial conditions on various combinations of networks have been also explored as well, and no significant changes are observed.

\begin{figure}[b]
\includegraphics[angle=0,scale=.55]{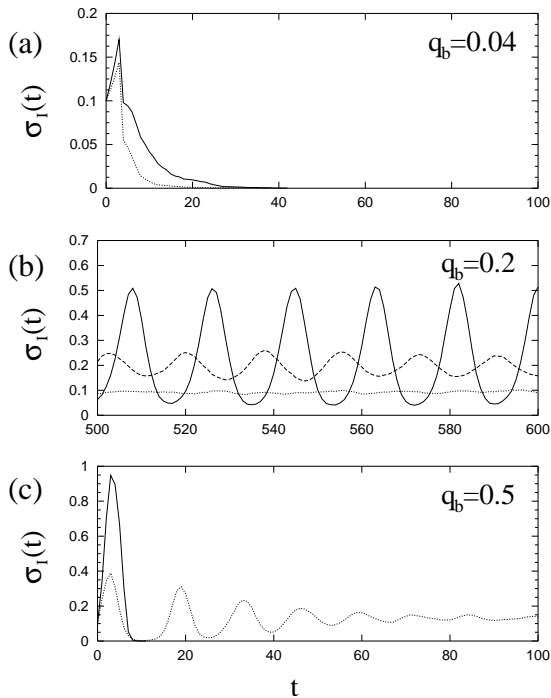}
\caption{The numerical simulations where the BL is described by small-world network and the HL is described by scale-free network. A solid line represents $q_{h}=0$ (no prevention), and a dotted line does $q_{h}=1$ (perfect prevention). (a) For small value of $q_{b}$, (b) for an intermediate range of $q_{b}$, where the dashed line is for the case of $q_{h}=0.04$, (c) for large value of $q_{b}$.}
\end{figure}

We have performed a simulation with various values of probabilities $q_{h}$ and $q_{b}$. The density of the infected nodes of the system, denoted by  $\sigma_{I}(t)$, is an important dynamical variable in this study. Figure 3 shows the evolution of $\sigma_{I}(t)$ for various values of $q_{h}$ and $q_{b}$. The results are classified into three categories depending on infection probability $q_{b}$. (1) For small values of $q_{b}$, it is found that $\sigma_{I}(t)$ vanishes in a finite time for all range of $q_{h}$ (Fig. 3(a)). Here it is found that immunization by prevention lowers the peak of $\sigma_{I}(t)$. (2) The results for an intermediate range of $q_{b}$ are shown in Fig. 3(b). When prevention is absent for this range, Fig. 3(b) shows that $\sigma_{I}(t)$ globally oscillates in the same way as was reported in Ref. \cite{KA2001}. When prevention is in effect, the amplitude of the global oscillation gets smaller as the prevention probability $q_{h}$ strengthens. (3) For large values of infection probability $q_{b}=0.5$, we observe a rather unexpected result. Figure 3(c) displays that prevention plays an adverse role in this case and actually helps the disease survive, rather than eradicate it.

\begin{figure}[b]
\includegraphics[angle=0,scale=.4]{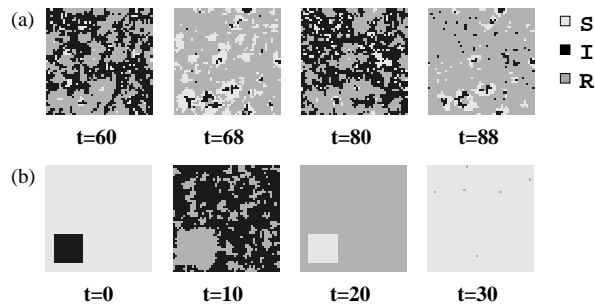}
\caption{Spatial characteristics showing the competition of two time scales. Underlying networks are 2-dimensional small-world networks with $N=50 \times 50$, $k=4$, and $p=0.1$. (a) The speed of spreading is relatively low, the oscillatory behavior establishes. $\tau_{I}=5$, $\tau_{R}=10$, and $q_{b}=0.35$. From the definitions of two time scales, $T_{d}\sim 15$ and $T_{s}\sim 4.66$. (b) The disease spreads over the system before initially infected nodes become refractory. $\tau_{I}=7$, $\tau_{R}=14$, and $q_{b}=0.5$, thus $T_{d}\sim 21$ and $T_{s}\sim 2.33$.}
\end{figure}

For a discussion, it is useful to introduce two characteristic time scales; one is the characteristic time scale of the disease ($T_{d} \sim \tau_{I} + \tau_{R}$) and the other is the time taken for the disease to sweep over the whole system ($T_{s}$), which is a function of $q_{b}$, $q_{h}$ and we write its simplest form as $L_{b}(\frac{1}{\tau_{I} q_{b}}+\tau_{R}q_{h})$, where $L_{b}$ is the characteristic path length of the BL. If $T_{d}$ is translated into the consumption of a species, $T_{s}$ would be the natural resources for the species. In case of $T_{s}\sim T_{d}$, let us assume that initially infected nodes are distributed locally, then the disease spreads over the whole system. As the almost infected nodes end their infection cycles and few susceptible nodes are left in some time, $\sigma_{I}(t)$ shows a steep collapse (Fig. 4(a)). This is the way the system exhibits oscillatory behavior, and its period is expected to be slightly longer than $T_{d}$. 
On the other hand, when $T_{s}$ is much smaller than $T_{d}$, then a disease soon cannot find any more $S$'s left and becomes exterminated. Infection occurs only when $I$-state node meets $S$-state node and is obstructed by $R$-state node. Since $q_{b}$ is quite large, the disease spreads very rapidly over the whole system, reaching $R$-state everywhere at some time, then we see that the disease cannot survive any more (Fig. 4(b)). Prevention, however, hinders this situation by always preserving $S$-state nodes, and thus makes the disease a chronic one in the end. This does resolve the paradoxical result in Fig. 3(c).

Figure 5 shows a phase diagram in the ($q_{h}$, $q_{b}$) plane, where darkened areas represent the regions of eradication of the disease, while white areas represent the regions of survival. In the lower dark region disease is eradicated because it takes too long time to prevail, i.e. $T_{s} \gg T_{d}$. On the contrary, it also does when it prevails too fast for individuals to recover themselves, as in the upper dark region where $T_{s} \ll T_{d}$. The white regions corresponds to the situation when $T_{s}\sim T_{d}$. For $q_{h}=0$, one can find the range  $T_{s0}<T_{s}< T_{s1}$ where the disease survives. The survival region in the plane of $(q_{b},q_{h})$ in Fig. 5 fits well to this range of $T_{s}$. For low values of $q_{h}$, prevention becomes significant for the regions where $q_{b}$ is high, but becomes insignificant in the region where $q_{b}$ is low. Some analytic explanation would be helpful. Assuming that the system is continuous in time and each $S$-state node with $k$ neighbors has the probability to have $k \sigma_{I}$ infected neighbors, we propose the following three ordinary differential equations in general:
\begin{widetext}
\begin{equation}
\begin{split}
\frac{d\sigma_{S}(t)}{dt}&= f(\sigma_{R}; \tau_{R}) - \sigma_{S} \left[
\sum_{k=1}^{N-1} p_k \{ 1-(1-q_b)^{k\sigma_I} \} +
\sum_{k=1}^{N-1} p'_k \{ 1-(1-q_h)^{k\sigma_I} \} \right], \\
\frac{d\sigma_{I}(t)}{dt}& = \sigma_{S}\sum_{k=1}^{N-1} p_{k} \{ 
1-(1-q_b)^{k\sigma_I} \} - g(\sigma_{I}; \tau_{I}), \\
\frac{d\sigma_{R}(t)}{dt}& = \sigma_{S}\sum_{k=1}^{N-1} p'_{k} \{ 
1-(1-q_h)^{k\sigma_I} \} + g(\sigma_{I}; \tau_{I})
-f(\sigma_{R}; \tau_{R}),
\end{split}
\end{equation}
\end{widetext}

where $\sigma_{S}$, $\sigma_{I}$ and $\sigma_{R}$ represent the densities of susceptible, infected and refractory nodes, respectively. $p_{k}$ and $p'_{k}$ are the normalized connectivity distributions of BL and HL networks, and $f$ and $g$ are monotonically increasing functions of $\sigma_{R}$ and $\sigma_{I}$ respectively, satisfying $f(0; \tau_{R})=g(0; \tau_{I})=0$. $\tau_{R}$ and $\tau_{I}$ mean relaxation times. Note that this set of equations already satisfies $\sigma_S(t)+\sigma_I(t)+\sigma_R(t)=1$.

\begin{figure}[b]
\includegraphics[angle=0,scale=.4]{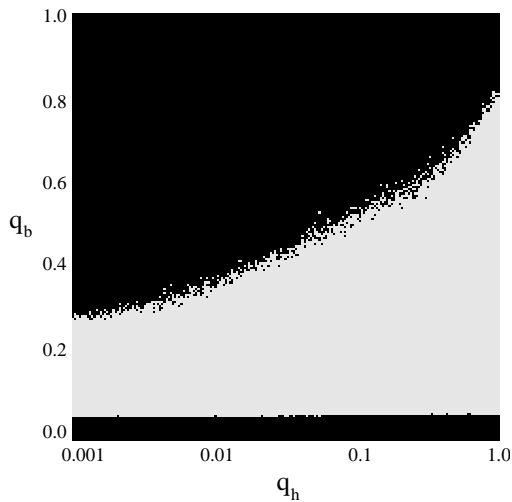}
\caption{The phase diagram which shows the regions of eradication (dark) and survival (clear) activity of the system in the ($q_{h}$, $q_{b}$) plane. Each point is averaged over 100 realizations. For low value of $q_{b}$ the immunization by prevention is irrelevant. For low value of $q_{h}$ the effect of prevention appears clearly.}
\end{figure}

From the stationary condition, one can find a threshold value of $q_{b}$ as a function of the average connectivity $\langle k\rangle =\sum_{k} k p_k$ and the infection period $\tau_I $, above which the disease survives. Assuming $g(\sigma_{I};\tau_{I})=\frac{\sigma_{I}}{\tau_{I}}$, in the limits of $\sigma_{S} \rightarrow 1$ and $\sigma_{I},\ \sigma_{R} \rightarrow 0$, we obtain the following analytic results for the threshold value of $q_{b}$,
\begin{equation}
q_{b,th}=1-e^{-1/\langle k\rangle \tau_I}.
\end{equation}
The fact that $q_{b,th}$ is not a function of $q_{h}$ explains why immunization by prevention seems almost irrelevant in the low $q_{b}$ region. In Fig. 6, the analytic results are compared with the numerical calculations. It becomes clear now that the shorter the infection period, the larger values of $q_{b,th}$ are needed for the disease to survive. 

\begin{figure}[b]
\includegraphics[angle=0,scale=.4]{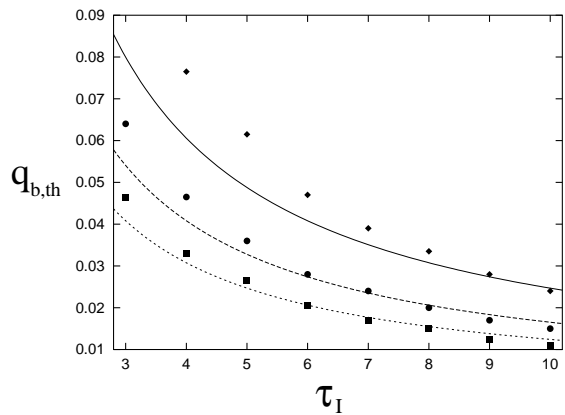}
\caption{Threshold values above which the disease survives. The curves show the equations for $q_{b,th}$ as a function of $\tau_{I}$. The diamonds, circles and squares show the $q_{b,th}$ in the case of $\langle k\rangle=4, 6, 8$ respectively by numerical results.}
\end{figure}

\begin{figure}
\includegraphics[angle=0,scale=.4]{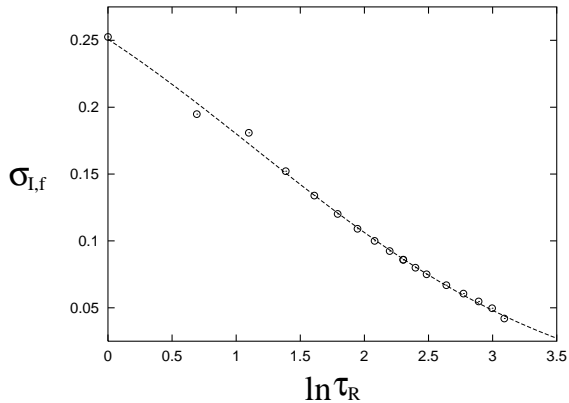}
\caption{The levels of small fluctuation $\sigma_{I,f}$ with perfect prevention, $q_{h}=1$, in Fig. 3(b), as a function of the recovery period $\tau_{R}$. The relation shows the Boltzmann sigmoidal behavior. The dots represent the numerical results and the line does the sigmoidal fitting.}
\end{figure}

Lastly we investigate the effect of variation of $\tau_{R}$ on the levels of small fluctuation $\sigma_{I,f}$ in case of perfect prevention, $q_{h}=1$, in Fig. 3(b). For larger $\tau_{R}$ the levels of fluctuation is expected to lower, which coincides with the numerical simulations. Figure 7 shows the Boltzmann sigmoidal behavior of the relation: $\sigma_{I,f}= \frac{A}{\tau_{R}^{\alpha} +B}+C$, where $A$, $B$, and $C$ are the positive constants, and $\alpha\approx 0.77$ from the fitting. For large $\tau_{R}$ the equation approximates to the power law with the exponent $-\alpha$. We can explain the above result from Eq. (2). In case of $q_{h}=1$ three variables have the nonzero fixed values which are $\sigma_{S,f}$, $\sigma_{I,f}$, and $\sigma_{R,f}$. From the stationary condition, assuming $g(\sigma_{I};\tau_{I})=\frac{\sigma_{I}}{\tau_{I}}$ again, $\sigma_{I,f}=(f(\sigma_{R,f};\tau_{R})-\sigma_{S,f})\tau_{I}$ is obtained. $f(\sigma_{R,f};\tau_{R})\sim \frac{1}{\tau_{R}}$ since $\tau_{R}$ means the relaxation time. Even for the infinite recovery period the disease does not eradicate because $C \not= 0$.

To summarize, we studied the immunization dynamics in epidemics by introducing a 2-layer network model, where an epidemic spreads on a small-world network of the system while the dynamical immunization by prevention takes place on a scale-free network of the same system. The prevention process indeed turns the global periodic rages of an epidemic into a small fluctuation. It is also found that prevention may help the disease survive when certain conditions are met. We introduced two time scales, $T_{s}$, $T_{d}$, and explained the numerical results. Recalling that we did not include any genetic processes here, such as occurrence of immunity-resistant disease through mutation, all of these should be purely dynamical effects. 

\begin{acknowledgments}

We thank H. Jeong for helpful comments. This work was supported by grant No. R01-1999-000-00019-0 from the Basic Research Program of the Korea Science and Engineering Foundation and also by grant No. KRF-2000-015-DP0097 from the Korea Research Foundation.

\end{acknowledgments}


\begin{thebibliography}{00}
\bibitem{Ls} The characteristic path lengths of scale-free and small-world networks are $L_{SF}(m_0=3)\approx 4.30$, $L_{SW}(p=0.5)\approx 5.73$.
\end{thebibliography}
\end{document}